# Reduction Of Spin Injection Efficiency by Interface Spin Scattering

R. M. Stroud, A. T. Hanbicki,[a] Y. D. Park,[a,b] A. G. Petukhov[c] and B. T. Jonker
*Naval Research Laboratory*, Washington, DC 20375

G. Itskos, G. Kioseoglou, M. Furis and A. Petrou
*State University of New York at Buffalo*, Buffalo, NY 14260

Abstract

We report the first experimental demonstration that interface microstructure limits diffusive electrical spin injection efficiency across heteroepitaxial interfaces. A theoretical treatment shows that the suppression of spin injection due to interface defects follows directly from the contribution of the defect potential to the spin-orbit interaction, resulting in enhanced spin-flip scattering. An inverse correlation between spin-polarized electron injection efficiency and interface defect density is demonstrated for ZnMnSe/AlGaAs-GaAs spin-LEDs with spin injection efficiencies of 0 to 85%.

[a] National Research Council Postdoctoral Fellow at NRL
[b] Current address: School of Physics, Seoul National University, Seoul, Korea
[c] Permanent address: Physics Dept., South Dakota School of Mines and Technology, Rapid City, South Dakota 57701

Electrical injection of spin polarized carriers into a semiconductor across a heteroepitaxial interface has now been demonstrated for a variety of all-semiconductor systems, with directly measured injection efficiencies greater than 50 % [1-3]. In contrast, the measured effects for spin-polarized electron injection in metal-semiconductor systems are only ~0.1-1% [4] and some argue that these results may be complicated by contributions not related to spin injection [5]. The reasons for the dramatically larger effects observed for the all-semiconductor systems include better conductivity and band matching, and the ability to grow isostructural interfaces. However, the actual extent to which physical properties such as interface microstructure limit polarized injection efficiency, and ultimately spintronic device performance, is not well known. One theoretical calculation [6] has shown that the broken symmetry of bonding across a perfectly ordered interface in itself reduces the injection polarization. Less subtle effects at the interface, such as strain, dislocations, or roughness, could result in a more pronounced reduction in spin polarization.

We have studied the effects of the interface microstructure on spin polarized injection efficiency using the $Zn_{1-x}Mn_xSe/Al_yGa_{1-y}As$-GaAs spin-LED as a test system (Fig. 1). In this device [7], spin polarized electrons are injected from the ZnMnSe into the AlGaAs-GaAs quantum well, where radiative recombination of the carriers results in the emission of circularly polarized light. Since the quantum selection rules mathematically relate the circular polarization of the light emitted along the surface normal to the spin polarization of the carriers, the spin-LED provides a quantitative, model-independent measure of spin injection efficiency [3]. Our results show that the spin injection efficiency correlates inversely with the density of defects nucleating at the ZnMnSe/AlGaAs interface. This correlation is explained by a model that incorporates spin-orbit (Elliot-Yafet) scattering, and shows that the asymmetric potential of the interface defect results in strong spin-flip scattering in the forward direction.

Samples were grown by molecular beam epitaxy (MBE) in a multichamber system. The growth was initiated with a 1 um p-type GaAs buffer layer on semi-insulating GaAs (001) substrates. On top of the buffer, the LED structure was grown (50 nm p-doped $Al_{0.9}Ga_{0.1}As$, 15 nm GaAs quantum well, 50 nm n-doped $Al_{0.9}Ga_{0.1}As$), followed by the n-$Zn_{0.94}Mn_{0.06}Se$ injector, which was capped in some cases by an $n^+$ZnSe contact layer. The use of a 15 nm GaAs quantum well lifts the heavy hole-light hole band degeneracy [3]. The density of defects at the ZnMnSe/AlGaAs interface was crudely controlled by varying the net thickness of the II-VI layers.

The samples were patterned into surface-emitting LEDs 200 to 400 microns in diameter using standard photolithography and chemical etching techniques. The spin injection efficiency was determined by measuring the circular polarization of the GaAs heavy hole free exciton in the Faraday geometry using a quarter wave plate followed by a linear polarizer and spectrometer. In this configuration, the circular polarization is equal to the electron spin polarization in the quantum well, and thus provides a direct measure of spin injection efficiency from the ZnMnSe contact. Further details of the growth, device fabrication and polarization analysis may be found elsewhere [3].

Cross-sectional samples in (110) orientations for transmission electron microscopy (TEM) studies were prepared by hand grinding and argon-ion milling. Bright-field / dark-field image pairs were collected in order to reveal the defect type and concentration, using a Philips CM30 300 kV TEM. High-resolution images of the interface microstructures were obtained using a Hitachi H9000 300 kV TEM.

The most prevalent defects observed were stacking faults (SFs) in <111> directions, nucleating at or near the ZnMnSe-AlGaAs interface. No secondary phases, Mn clusters, or defects in the AlGaAs-GaAs LEDs were found. The SFs were easily observed in dark-field images (g = $<2\bar{2}0>$) as diagonal lines extending from the heterointerface to the ZnMnSe film surface (Fig. 2). The number density of these SFs, calculated by counting the number of SFs observed for a series of dark field images and dividing by the total cross-sectional area, was found to vary from 2 x $10^8$/cm$^2$ to $10^{10}$/cm$^2$, with an estimated uncertainty of a factor of two. These SFs are a well-known problem in ZnSe/GaAs epitaxy, typically occurring at concentrations of $10^8$-$10^9$/cm$^2$ [8], and are attributed to both lattice strain [9] and the formation of Se dimers at the interface [10]. Away from the immediate vicinity of a SF, high-resolution lattice images (Fig. 3) reveal a structurally well-ordered interface.

The spin injection efficiency of the ZnMnSe/AlGaAs-GaAs spin-LEDs correlates inversely with the observed SF density, as shown in Figure 4. This correlation clearly demonstrates a link between the injection efficiency and interface microstructure. It is remarkable to note that polarized injection efficiencies as high as 85% can be realized across the II-VI / III-V heterointerface despite moderately high ($10^8$ - $10^9$ cm$^{-2}$) SF densities, attesting to the robust nature of the spin injection process. A least squares fit to the data (dashed line in Fig. 4), shows a decrease in injection efficiency of a only factor of five for a 100-fold increase in SF density. Complete suppression of the spin polarized injection was only achieved for a sample in which the AlGaAs surface was exposed to air and then chemically treated using an ammonium sulfide surface passivation procedure [11] prior to the growth of the ZnMnSe, resulting in a very rough interface in addition to the highest SF density of $10^{10}$ cm$^{-2}$.

To theoretically analyze the relationship between spin polarization and defects, we have adapted a model of the scattering properties of low symmetry defects that includes spin-orbit interactions. In principle, spin-flip scattering in the ZnMnSe can occur anywhere along the extended planar SFs. However, Oestreich *et al.* [12] have noted that in a semimagnetic semiconductor, the minority spin lifetime is exceedingly short on the time scale of diffusive transport when the conduction band edge is split by a sufficiently large magnetic field. If a spin-flip scattering event occurs far above the ZnMnSe/AlGaAs interface, the carrier spin quickly relaxes to the majority spin channel before reaching the interface and being injected. Therefore, the relevant spin scattering in the spin-LED for diffusive transport occurs at the interface. Thus, we need only consider scattering from the line defect associated with the intersection of each SF plane with the interface plane.

The spin-orbit Hamiltonian associated with the potential of a lattice defect can be written from Kane's model [13] as:

$$H_{SO} = \frac{\hbar \Delta}{3mE_g^2} \vec{\sigma} \left[ \vec{\nabla} U(\vec{r}) \times \vec{p} \right] \qquad (1)$$

where $\Delta$ is the spin-orbit splitting in the valence band, $E_g$ is the band gap, $m$ is the effective mass, $\vec{p}$ is the momentum operator, and $U(\vec{r})$ is the potential of the defect. The Hamiltonian (1) is responsible for the Elliot-Yaffet spin-scattering mechanism in bulk III-V or II-VI semiconductors [13]. The potential for a linear defect with radial symmetry, i.e., the line defect associated with the intersection of the stacking fault with the interface plane described above, can be written [14]:

$$U(\vec{r}) = (A/\rho) f(\rho) \sin \varphi, \qquad (2)$$

where the conventional polar coordinate system is chosen with the $(1\bar{1}0)$-axis as the defect axis (see Figure 5). This potential describes a line dipole along $(1\bar{1}0)$ with a dipole moment parallel to the (110) axis. The distance from a probe point to the axis of the linear defect is $\rho$, $\varphi$ is the azimuthal angle, $f(\rho)$ is a Thomas-Fermi screening function such that $f(0) = 1$ and $f(\infty) = 0$, and $A$ is a constant combining several material parameters such as deformation potential, dielectric constant, and the Poisson ratio [14].

Plane-wave matrix elements of $U$ and $H_{SO}$ determine the scattering matrix, which can be written in the Born approximation as follows:

$$S(\vec{k},\vec{k'}) = i(\langle \vec{k} | U | \vec{k'} \rangle + \langle \vec{k} | H_{SO} | \vec{k'} \rangle) \equiv g(\vec{k},\vec{k'}) + ih(\vec{k},\vec{k'})\vec{\sigma}\hat{n}(\vec{k},\vec{k'}). \quad (3)$$

Here $g(\vec{k},\vec{k'})$ and $h(\vec{k},\vec{k'})$ are the non-spin-flip and spin-flip scattering amplitudes respectively [15,16], and vector $\hat{n}(\vec{k},\vec{k'})$ is a unit vector normal to the incident and scattered wave vectors.

The spin polarization can be calculated straightforwardly from the scattering matrix following Merzbacher [15], and Landau and Lifschitz [16]. We are primarily concerned with electrons moving towards the GaAs quantum well (k // $(00\bar{1})$, and having a large longitudinal component of the initial spin polarization $\vec{P}_0(\vec{k}) \parallel \vec{k}$, corresponding to spin injection across the interface with an out-of-plane applied magnetic field. Calculating the final spin polarization $\vec{P}(\vec{k'})$ of the electron scattered in the direction of $\vec{k'}$, we introduce the function

$$\pi(\vec{k},\vec{k'}) = \frac{1}{2} + \frac{\vec{k}\vec{P}(\vec{k'})}{2\vec{k}\vec{P}_0(\vec{k})} = \frac{1}{2} + \frac{|g(\vec{k},\vec{k'})|^2 - |h(\vec{k},\vec{k'})|^2 + 2|h(\vec{k},\vec{k'})|^2 \vec{k}\hat{n}(\vec{k},\vec{k'})/k}{2\left(|g(\vec{k},\vec{k'})|^2 + |h(\vec{k},\vec{k'})|^2\right)}, \quad (4)$$

describing angular distribution of the spin polarization. This function is a natural measure of the probability of spin-flip processes. Its value lies between 0 and 1, with $\pi(\vec{k},\vec{k'}) = 0.5$ corresponding to zero spin polarization. When $\pi(\vec{k},\vec{k'})$ is close to zero the spin-flip processes dominate and result in significant degradation of the spin polarization.

The angular distribution of the spin polarization, $\pi(\vec{k},\vec{k'})$, about the defect axis is presented in Fig. 5 (solid lines) for different values of the dimensionless parameter, $kr_0$, where $r_0$ is the Thomas-Fermi screening distance ($kr_0 \sim 1$ corresponds to the electrons at the Fermi level). The outer dashed circle corresponds to 100% spin polarization (no spin-flip scattering), while the inner dashed circle corresponds to zero spin polarization, where one half of the carriers have flipped their spin due to interaction with the defect potential. The spin-flip effect is quite pronounced, and results in a significant reduction of the spin polarization in forward scattering. Thus the longitudinal component of the spin polarization of the electrons scattered forward at small angles (towards the GaAs quantum well) will be significantly reduced as a result of scattering. If one makes the simple assumption that an electron will "cross" the interface with equal probability at any interface site, and that a fraction $n$ of the interface sites are defects, then the linear dependence of spin injection efficiency on defect density observed experimentally (Figure 4) follows directly. Figure 5 also clearly shows that the effects of spin depolarization become dominant for hot electrons with $kr_0 \gg 1$ for all directions. Qualitatively similar results are obtained for other orientations of the dipole moment. This angular distribution and energy dependence are clearly important factors to consider when attempting to measure spin polarization in any given structure.

In summary, the high spin injection efficiency of the all–semiconductor ZnMnSe/AlGaAs-GaAs spin-LEDs has enabled the first demonstration of a direct correlation between the spin injection efficiency across a heteroepitaxial interface and the structural quality of that interface. An increase by two orders of magnitude in the density of interface-nucleated stacking faults produced a decrease in the spin injection efficiency by a factor of five. The polarization reduction in the presence of interface defects with non-spherical symmetry arises naturally from the contribution of the asymmetric defect potential to the spin-orbit orbit coupling. Because interface defects are generic to heteroepitaxial systems, these results have implications for all spin transport heterostructures.

This work was supported by the Office of Naval Research and the Defense Advanced Research Projects Agency *SpinS* program. YDP and AH gratefully acknowledge support from the NRL/NRC postdoctoral fellowship program. AGP gratefully acknowledges financial support from the ASEE-ONR Summer Faculty Program and NSF Grant No DMR-0071823. He also kindly acknowledges warm hospitality found at the Naval Research Laboratory during his summer visit.

**FIGURE CAPTIONS**

**Figure 1**. Schematic of the ZnMnSe/AlGaAs-GaAs spin-LED. Spin polarized electrons are injected across the ZnMnSe–AlGaAs interface into the GaAs quantum well. Radiative recombination in the QW results in circularly polarized light emission.

**Figure 2**. Dark-field cross-sectional transmission electron micrograph of a ZnMnSe spin-LED. Stacking faults in <111> directions are observed to nucleate at the ZnMnSe-AlGaAs interface. This sample had a stacking fault concentration of 3.5 x $10^9$ $cm^{-2}$ and spin-injection efficiency of 49%.

**Figure 3**. High-resolution lattice image of a ZnMnSe-AlGaAs-GaAs spin-LED. The ZnMnSe/AlGaAs interface (black line) is structurally well-ordered in areas between interface-nucleated stacking faults.

**Figure 4**. Correlation of spin injection efficiency with stacking fault density. The dashed line is a least squares fit to the data.

**Figure 5**. Angular distributions of the spin polarization about the axis of the linear interface defect (the (-110) direction for different values of $kr_0$ following scattering from the defect. The outer dashed circle corresponds to 100% spin polarization, while the inner circle divides the regions of the negative and positive polarizations. The dipole moment is oriented along the (110) direction, i.e. in the interface plane. Spin-flip scattering

dominates for all electrons in the forward direction (towards the GaAs quantum well).

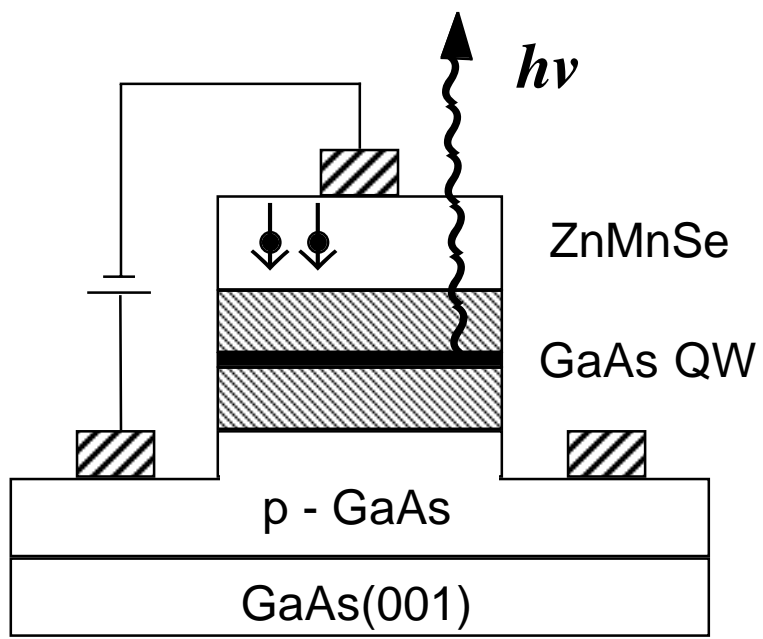

Fig. 1 – Stroud, *et al.*

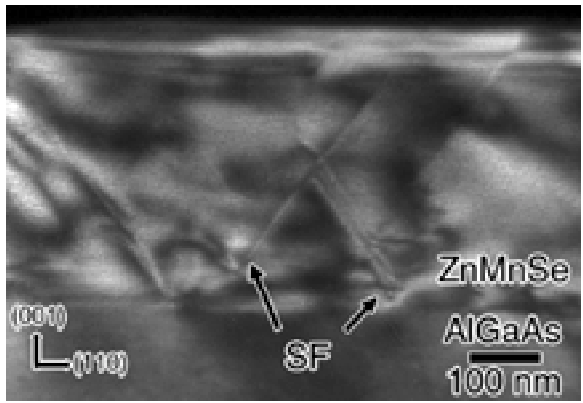

Figure 2 – Stroud, *et al.*

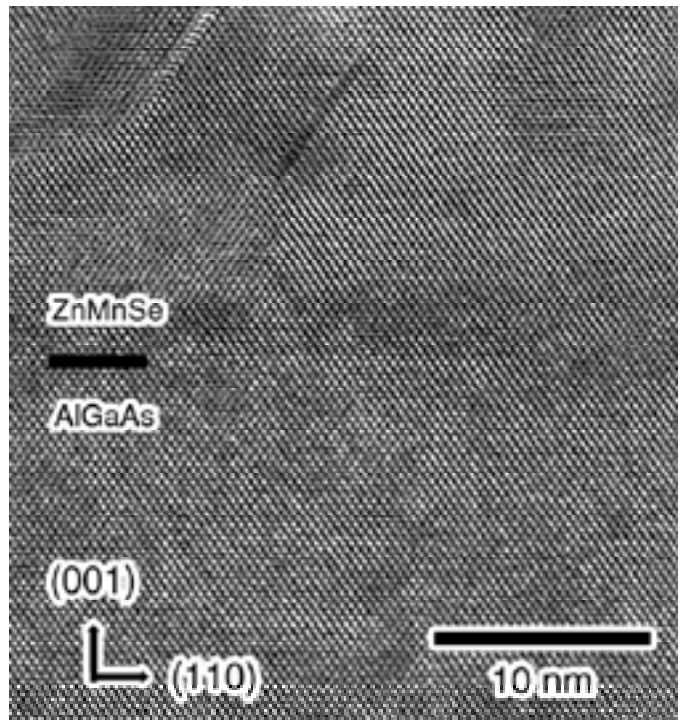

Figure 3 – Stroud, *et al.*

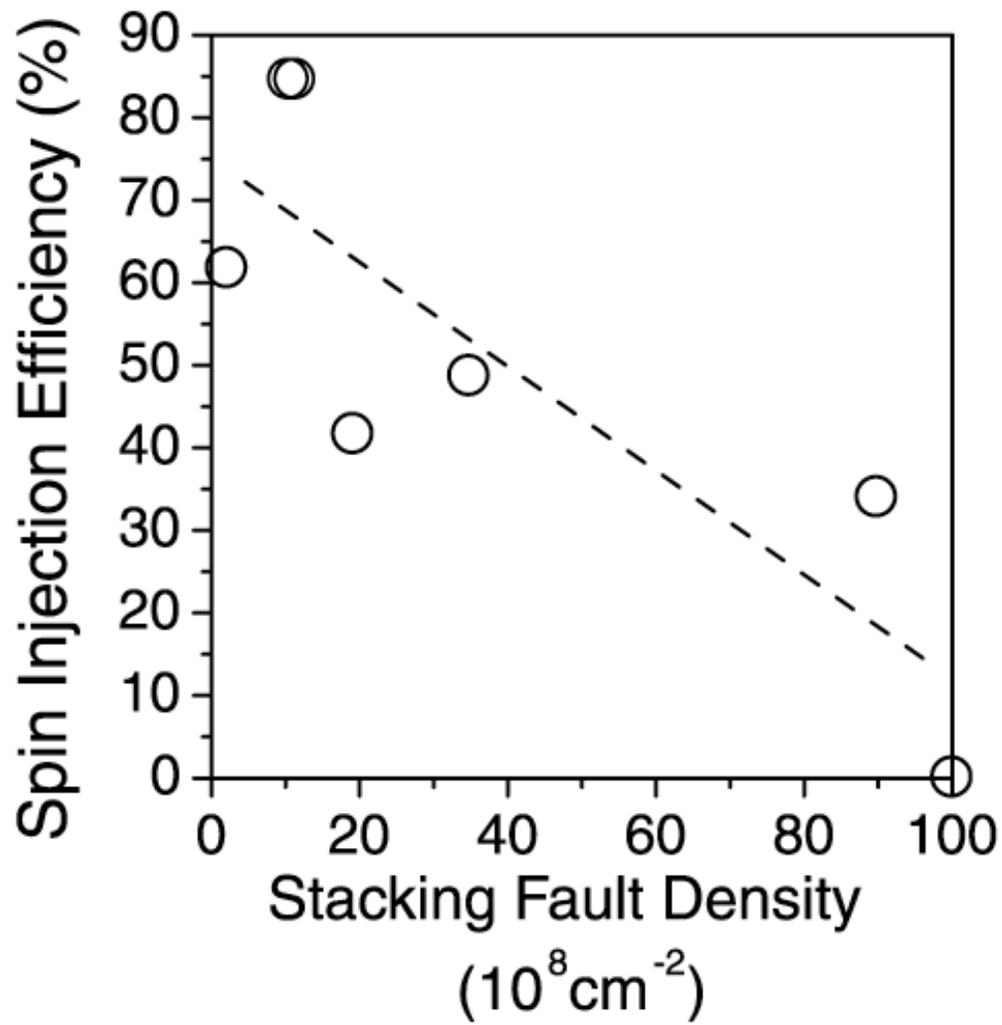

Figure 4 – Stroud, *et al.*

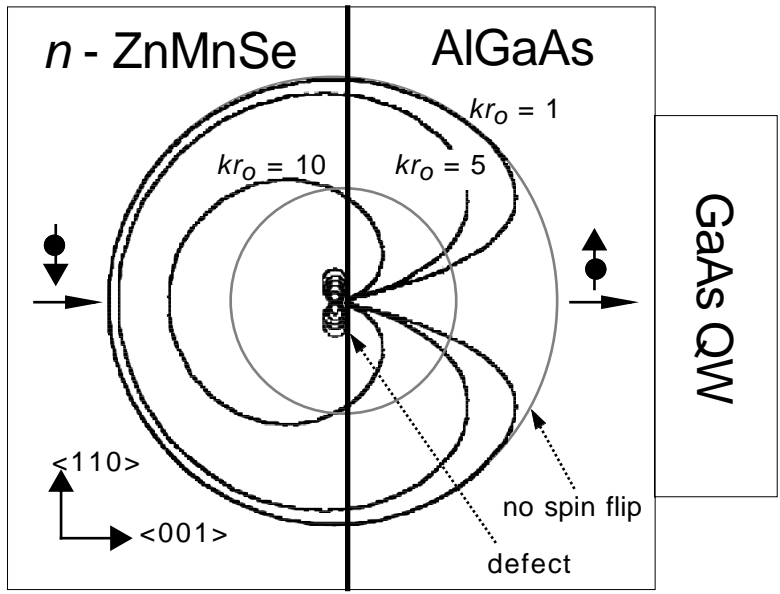

Figure 5 – Stroud, *et al.*